\documentclass[cameraready]{Interspeech}

\title{CTC-TTS: LLM-Based Dual-Streaming Text-to-Speech with CTC Alignment}

\author[affiliation={1}, orcid=0009-0005-8285-6615]{Hanwen}{Liu}
\author[affiliation={1}, orcid=0009-0003-1947-8362]{Saierdaer}{Yusuyin}
\author[affiliation={1}, orcid=0000-0001-6604-0951]{Hao}{Huang}
\author[affiliation={2}, orcid=0000-0002-9018-5074, correspondingauthor]{Zhijian}{Ou}

\address{
    $^1$ School of Computer Science and Technology, Xinjiang University, China \\
    $^2$ Speech Processing and Machine Intelligence (SPMI) Lab, Tsinghua University, China
}

\email{huanghao@xju.edu.cn, ozj@tsinghua.edu.cn}


\usepackage{multirow}
\usepackage{booktabs}
\usepackage{url}

\usepackage{stfloats}

\DeclareMathOperator*{\argmax}{argmax}
\newcommand{\tok}[1]{\ensuremath{\langle\text{#1}\rangle}}

\begin{document}

\maketitle

\begin{abstract}
    Large-language-model (LLM)-based text-to-speech (TTS) systems can generate natural speech, but most are not designed for low-latency dual-streaming synthesis. High-quality dual-streaming TTS depends on accurate text--speech alignment and well-designed training sequences that balance synthesis quality and latency. Prior work often relies on GMM-HMM based forced-alignment toolkits (e.g., MFA), which are pipeline-heavy and less flexible than neural aligners; fixed-ratio interleaving of text and speech tokens struggles to capture text--speech alignment regularities. We propose CTC-TTS, which replaces MFA with a CTC based aligner and introduces a bi-word based interleaving strategy. Two variants are designed: CTC-TTS-L (token concatenation along the sequence length) for higher quality and CTC-TTS-F (feature-dimension embedding stacking) for lower latency. Experiments show that CTC-TTS outperforms fixed-ratio interleaving and MFA-based baselines on streaming synthesis and zero-shot tasks.
\end{abstract}

\section{Introduction}

In recent years, large language model (LLM)-based methods \cite{valle, seedtts, cosyvoice, cosyvoice2} that treat text-to-speech (TTS) as a language modeling task have gained widespread attention. These methods typically convert continuous speech signals into discrete token sequences with a neural audio codec (NAC) \cite{wavtokenizer, soundstream, encodec}, predict the sequence using an LLM, and finally reconstruct the speech waveform via the NAC decoder.

However, most LLM-based TTS systems are not designed for low-latency dual-streaming synthesis (simultaneous text input and speech output). \cite{valle, seedtts, cosyvoice}. High-quality dual-streaming TTS requires accurate text--speech alignment and well-designed training sequences that balance synthesis quality and latency. Prior work often relies on GMM-HMM based forced-alignment toolkits (e.g., Montreal Forced Aligner, MFA) \cite{mfa}, which are pipeline-heavy and less flexible than neural aligners. Meanwhile, fixed-ratio interleaving of text and speech tokens struggles to capture alignment regularities between text and speech, making it more difficult for the model to learn reliable temporal dependencies. Alignment-aware interleaving can alleviate this issue but still typically depends on MFA; existing methods adopt different alignment units such as words, Byte Pair Encodings (BPEs) \cite{bpe}, and phonemes and diverse sequence organizations, as reviewed in Section~\ref{sec:related}.

To address these limitations, we propose \textbf{CTC-TTS}\footnote{Code is available at \url{https://github.com/thu-spmi/CTC-TTS}.}, a Connectionist Temporal Classification (CTC)-based \cite{ctc} dual-streaming TTS method that improves both alignment and sequence organization. CTC alignment introduces a blank symbol and yields a robust structural correspondence without requiring frame-accurate phoneme boundaries. This level of alignment is sufficient for an autoregressive model to learn to map local phoneme groups to speech tokens, reducing learning complexity compared with fixed-ratio interleaving and avoiding the heavy pipeline of MFA. We construct \emph{bi-word} blocks: phonemes of the current word plus the separator and phonemes of the next word, followed by the speech tokens aligned to the current word.

To balance synthesis quality and latency, we design two variants based on different interleaving implementations of phonemes and speech tokens for bi-words. \textbf{CTC-TTS-L} concatenates text and speech tokens along the sequence length dimension to enhance generation quality. \textbf{CTC-TTS-F} stacks text and speech embeddings along the feature dimension, enabling synthesis to start from the first phoneme and thus reducing first-packet latency. Experiments show that CTC-TTS outperforms strong baselines on streaming synthesis and zero-shot tasks.

Our main contributions are: (1) a lightweight CTC-based phoneme-speech alignment procedure for LLM-based TTS; (2) a bi-word interleaving strategy with compact look-ahead; (3) two variants (CTC-TTS-L/F) that realize different quality-latency trade-offs and improve streaming and zero-shot results.

\section{Related Work}\label{sec:related}

Early LLM-based TTS methods concatenate full speech tokens after the complete text input, leading to large first-packet latency (FPL) \cite{valle, cosyvoice}. 
Dual-streaming methods therefore construct interleaved text--speech sequences so that an autoregressive decoder can generate speech tokens while text tokens arrives.

\textbf{Non-aligned interleaving.} IST-LM \cite{istlm}, CosyVoice2 \cite{cosyvoice2}, and StreamMel \cite{streammel} use fixed-ratio interleaving of text and speech. LLMVox \cite{llmvox} stacks phoneme and speech embeddings in the feature dimension to reduce FPL.

\textbf{Aligned interleaving.} SpeakStream \cite{speakstream} uses word-level forced alignment to pair chunked speech with a window of text words. SyncSpeech \cite{syncspeech} applies MFA for speech--BPE \cite{bpe} alignment and generates speech tokens per BPE token. ELLA-V \cite{ellav} employs MFA for phoneme--speech alignment and aligned interleaving, outperforming the non-aligned VALL-E \cite{valle}.

\textbf{Interleaving design.} Interleaving schemes reflect quality-latency trade-offs in streaming speech synthesis, by using different amounts of look-ahead. 
SpeakStream trades latency for quality via the text-window size and speech hop. SyncSpeech introduces duration modeling, increasing model complexity. ELLA-V adopts global advance (prepending all phonemes for global context) and local advance (shifting phonemes forward for local context). In this paper, we adopt a bi-word interleaving unit to balance synthesis quality and latency.

\section{CTC-TTS}
\subsection{Speech--phoneme alignment}\label{sec:align}
Given input speech features, a CTC-based automatic speech recognition (ASR) model outputs a posterior distribution over text labels\footnote{In this work, we use phonemes as text labels; exploring graphemes or subwords as alignment units is left for future work.} at each frame. From this distribution, we obtain the maximum-probability alignment path $\boldsymbol{\pi}^{*}$ via the Viterbi algorithm \cite{ctc}:
\begin{align}
\boldsymbol{\pi}^{*} = \argmax_{\boldsymbol{\pi}} \prod_{t=1}^{T} P(\pi_{t}\mid x_{t})
\label{equation:path}
\end{align}
where $x_{t}$ denotes the acoustic feature at frame $t$, $\boldsymbol{\pi}=(\pi_{1},\ldots,\pi_{T})$ is a frame-level alignment path over an extended alphabet (phonemes plus the blank symbol), and $P(\pi_{t}\mid x_{t})$ is the posterior probability at frame $t$ computed from the logits produced by the neural-network based acoustic model. In CTC, multiple consecutive frames may share the same label and blank labels may appear between phoneme labels; the standard collapse operation (merge repeats and remove blanks) maps a path to the corresponding phoneme sequence \cite{ctc}.

In the path $\boldsymbol{\pi}^{*}$, phoneme labels usually lag their actual speech onset, and blank labels may correspond to silence or adjacent phoneme regions \cite{ctc_align}. We therefore assign each blank label to its first subsequent phoneme. If $\boldsymbol{\pi}^{*}$ ends with blank labels, these trailing blanks are assigned to the final phoneme label, yielding the refined path $\hat{\boldsymbol{\pi}}$. In our experiments, the CTC model converts speech into acoustic features at 25 frames per second, while the NAC converts speech into a discrete token sequence $\mathbf{s}$ at 75 frames per second, the length ratio between $\hat{\boldsymbol{\pi}}$ and $\mathbf{s}=(s_{1},\ldots,s_{3 T})$ is 1:3. We thus map each phoneme label in $\hat{\boldsymbol{\pi}}$ to three corresponding speech tokens in $\mathbf{s}$, such that the $i$-th phoneme $\hat{\boldsymbol{\pi}}_{i}$ is aligned to the discrete speech tokens $[{s}_{3i-2},{s}_{3i-1},{s}_{3i}]$.

Unlike GMM-HMM forced alignment, the CTC alignment used here is not intended to provide frame-accurate phoneme boundaries. Instead, it provides a stable structural alignment that is sufficient for constructing word-level phoneme--speech blocks and for learning the mapping from local phoneme groups to speech tokens.

\subsection{Interleaving schemes}\label{sec:interleave}

\begin{figure}[t]
  \centering
  \includegraphics[width=\linewidth]{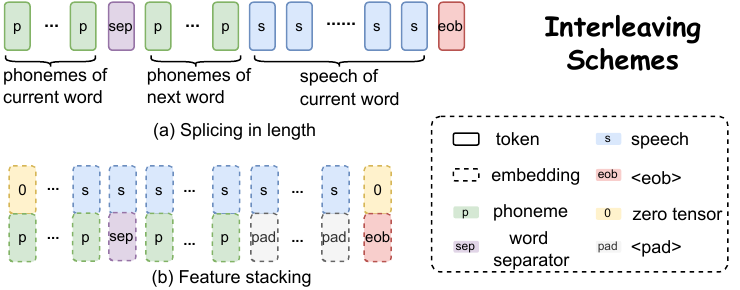}
  \caption{One bi-word block in the two text--speech interleaving schemes: (a) CTC-TTS-L and (b) CTC-TTS-F.}
  \label{fig:interleaving_schemes}
\end{figure}

After aligning phonemes and speech, we derive the speech sequence corresponding to the phonemes of each word. Since a word's pronunciation depends on its context, we construct the sequence illustrated in Figure~\ref{fig:interleaving_schemes}(a): after the phonemes of the current word, we append the word separator between the current and next word (space, comma, period, question mark and exclamation mark), together with the phonemes of the next word, followed by the speech tokens corresponding to the current word, and end with a block terminator \tok{eob}. For the $k$-th word, we refer to this as a (bi-word) phoneme--speech block $\mathbf{b}_{k}$. For the last word, which has no subsequent word, we insert an \tok{eos} as a placeholder for the future word. Concatenating all phoneme--speech blocks yields the complete training sequence for an utterance: $\mathbf{Y}=\mathbf{b}_{1}\oplus \mathbf{b}_{2}\oplus\cdots$. The model trained on such text--speech sequences constructed by length-wise concatenation is denoted as \textbf{CTC-TTS-L} (Figure~\ref{fig:interleaving_schemes}a).

CTC-TTS-L only starts generating speech tokens after receiving phonemes for the first two words, which increases first-packet latency. To mitigate this issue, we follow LLMVox \cite{llmvox} and instead stack phoneme and speech embeddings along the feature dimension (Figure~\ref{fig:interleaving_schemes}b). 
For each block, we append \tok{pad} tokens to the phoneme sequence so that its length matches the speech-token sequence length, then stack them along the feature dimension.
For initialization, the first phoneme of the initial block and the \tok{eob} token are both stacked with an all-zero tensor.
During inference, the model can generate speech tokens immediately after reading the first phoneme (stacked with zeros), which reduces first-packet latency. The model trained with feature-level stacking is denoted as \textbf{CTC-TTS-F}.

\subsection{Model architecture}\label{sec:arch}
\begin{figure*}[htbp]
  \centering
  \includegraphics[width=\linewidth]{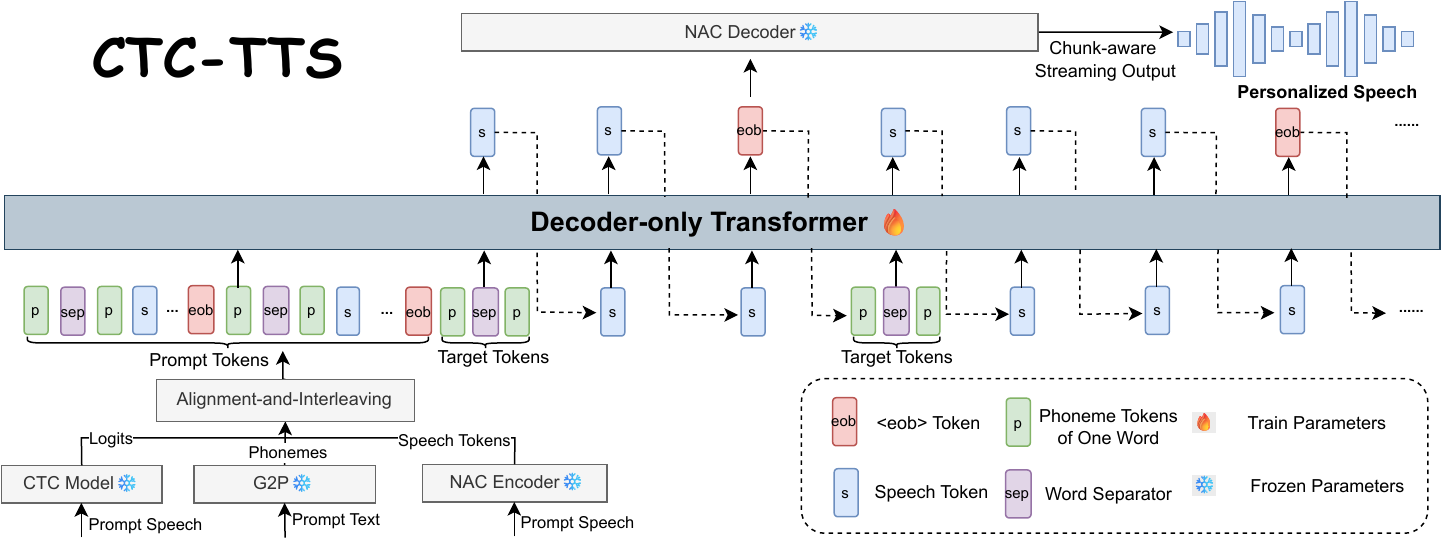}
  \caption{Overview of CTC-TTS-L. Components include: (1) a G2P model that converts text to phonemes; (2) a CTC-based ASR model for speech--phoneme alignment; (3) a decoder-only LM that models interleaved text and speech tokens; (4) a neural audio codec; and (5) an alignment-and-interleaving module implementing Sections~\ref{sec:align}--\ref{sec:interleave}. CTC-TTS-F shares the same components but uses feature-level stacking (Fig.~\ref{fig:interleaving_schemes}b) and is omitted for brevity. No prompts are required in single-speaker settings.}
  \label{fig:ctc_tts}
\end{figure*}

Figure~\ref{fig:ctc_tts} illustrates the CTC-TTS framework (shown for CTC-TTS-L); CTC-TTS-F is analogous and omitted for brevity. Since this paper adopts a single-codebook NAC, only one autoregressive Transformer is used to predict speech tokens and the $\tok{eob}$ symbol. Let the training sequence be rewritten as $\mathbf{Y}=y_{1},\cdots,y_{|\mathbf{Y}|}$, where $|\mathbf{Y}|$ denotes the sequence length. The training objective is to minimize the cross-entropy loss over speech tokens and $\tok{eob}$:

\begin{align}
\mathcal{L} = \sum_{t=1,\cdots,|\mathbf{Y}|;~{t} \notin \mathcal{T}} - \log P(y_{t}|y_{<t};\theta)
\label{equation:loss}
\end{align}
where $\mathcal{T}$ denotes the positions of text tokens (phonemes and $\tok{eos}$) in the interleaved sequence, and $\theta$ denotes the model parameters.

For CTC-TTS-L, the model starts generating speech tokens after reading the phonemes of the current word and the next word. It keeps generating until emitting $\tok{eob}$, which indicates the end of the current word's speech block. Inference terminates after $\tok{eob}$ is emitted for the last word.

For CTC-TTS-F, inference begins by receiving the first phoneme. Its embedding is stacked with an all-zero tensor along the feature dimension and fed into the model to generate one speech token. After each generated speech token, its embedding is stacked with the next phoneme embedding (or $\tok{pad}$ after the phonemes for two words are exhausted) and appended to the sequence. When $\tok{eob}$ is emitted, the current word's speech block ends; we then stack $\tok{eob}$ with zeros to generate the next speech token, and stack that token with the next word's first phoneme to start the subsequent block. Inference terminates when $\tok{eob}$ is emitted for the last word. Following LLMVox, we decode speech tokens to audio in chunks. For zero-shot dual-streaming synthesis, the process is similar but conditioned on pre-aligned speech and text prompts (prompt tokens in Figure~\ref{fig:ctc_tts}), which are typically reusable.

\section{Experimental Setup}
\subsection{Pre-trained models and dataset selection}

We use a LibriSpeech-trained monolingual Whistle \cite{librispeech,whistle} as the CTC model (115M parameters), which is a Conformer-based \cite{conformer} ASR model with weak phonetic supervision.
Phonetisaurus \cite{phonetisaurus}, a WFST-based \cite{wfst} G2P toolkit, converts English text to IPA phonemes. WavTokenizer \cite{wavtokenizer}, a single-codebook NAC, enables a single autoregressive model to predict all speech tokens. 
To obtain the speech–phoneme alignment, we use the \texttt{forced\_align} function in Torchaudio, which takes CTC posteriors as input and implements Eq.~(\ref{equation:path}) via the Viterbi algorithm.

\textbf{Single-speaker streaming experiments}. We evaluate dual-streaming generation with the two CTC-TTS variants against LLMVox \cite{llmvox}. Following LLMVox, we use the VoiceAssistant400K \cite{miniomni} single-speaker dataset (1750\,h training, 50\,h validation, 5\,h test) after filtering untranscribable utterances via Phonetisaurus. We reproduce LLMVox on this dataset for comparison and adopt the same chunk-aware streaming speech output paradigm as LLMVox.

\textbf{Multi-speaker zero-shot experiments}. We compare our alignment and interleaving schemes with baselines on two tasks (continuation and cross-speaker). Different model are trained on the 960-hour LibriSpeech dataset. For evaluation, the continuation task uses 4--10\,s utterances from \texttt{test-clean}, while the cross-speaker task uses Seed-TTS \cite{seedtts} \texttt{test-en}. We reproduce ELLA-V's local-advance setting with the same backbone, NAC, and data splits for fair comparison.

\begin{table}[th]
  \caption{Comparison between CTC-TTS and LLMVox. LLMVox interleaves text and speech at a fixed ratio and stacks embeddings in the feature dimension. CTC-TTS-F stacks embeddings in the feature dimension, while CTC-TTS-L concatenates tokens along the length dimension.  \#Params refers to model parameters, and NA indicates not applied.}
  \vspace{-0.2cm}
  \label{tab:ctc_llmvox_compare}
  \centering
  \small  
  \setlength{\tabcolsep}{1.5pt}
  \begin{tabular}{@{}c@{}c c c c c@{}} 
    \toprule
    \textbf{Method} & \textbf{\#Params} & \textbf{WER\%$\downarrow$} & \textbf{CER\%$\downarrow$} & \textbf{FPL-A$\downarrow$} & \textbf{UTMOS$\uparrow$} \\
    \midrule
    Ground Truth            & NA       & NA    & NA    & NA       & 4.27 \\
    LLMVox         & 31.5M   & 2.40  & 1.36  & 167     & \textbf{4.15} \\
    CTC-TTS-F      & 33.6M   & 1.80  & 1.04  & \textbf{159}     & \textbf{4.15} \\
    CTC-TTS-L      & 34.7M   & \textbf{1.50}  & \textbf{0.79}  & 210     & \textbf{4.15} \\
    \bottomrule
  \end{tabular}
   \vspace{-0.4cm}
\end{table}

\begin{table*}[th]
  \caption{Experimental results of different methods on the continuation task (\textbf{bold}: best, \underline{underline}: second best).}
    \vspace{-0.2cm}
  \label{tab:continuation}
  \centering
  \setlength{\tabcolsep}{5pt} 
  \begin{tabular}{c c c c c c c c c}  
    \toprule
    \textbf{Group} & \textbf{Method} & \textbf{\#Params} & \textbf{WER\%$\downarrow$} & \textbf{CER\%$\downarrow$} & \textbf{SPK$\uparrow$} & \textbf{UTMOS$\uparrow$} & \textbf{MOS$\uparrow$} & \textbf{SMOS$\uparrow$} \\
    \midrule
    Ground Truth & NA & NA & 1.92 & 0.69 & NA & 4.086 & $4.28\pm0.060$ & $4.60\pm0.048$ \\
    \midrule
    \multirow{2}{*}{Our Method} & CTC-TTS-F & 158.47M & 5.20 & 2.68 & \textbf{0.930} & 4.013 & \underline{$4.31\pm0.057$} & \underline{$4.58\pm0.050$} \\
    & CTC-TTS-L & 159.58M & \textbf{4.82} & \textbf{2.47} & \underline{0.929} & \textbf{4.050} & $\mathbf{4.33\pm0.061}$ & $\mathbf{4.60\pm0.049}$ \\
    \midrule
    \multirow{3}{*}{Ablation} & CTC+ELLA-V & 159.58M & 12.01 & 7.37 & 0.928 & \underline{4.021} & $4.00\pm0.062$ & $4.39\pm0.058$ \\
    & MFA+ELLA-V & 159.58M & 10.98 & 6.99 & 0.928 & \underline{4.021} & $3.94\pm0.066$ & $4.44\pm0.056$ \\
    & MFA+bi-word & 159.58M & \underline{5.14} & \underline{2.63} & \textbf{0.930} & 4.010 & $4.25\pm0.061$ & $4.50\pm0.051$ \\
    \bottomrule
  \end{tabular}
\end{table*}

\begin{table*}[th]
  \caption{Experimental results of different methods on the cross-speaker task.}
    \vspace{-0.2cm}
  \label{tab:cross}
  \centering
  \setlength{\tabcolsep}{5pt}  
\begin{tabular}{c c c c c c c c c}
\toprule
\textbf{Group} & \textbf{Method} & \textbf{\#Params} & \textbf{WER\%$\downarrow$} & \textbf{CER\%$\downarrow$} & \textbf{SPK$\uparrow$} & \textbf{UTMOS$\uparrow$} & \textbf{MOS$\uparrow$} & \textbf{SMOS$\uparrow$} \\
\midrule
Ground Truth & NA & NA & NA & NA & NA & 3.527 & $4.18\pm0.068$ & $4.14\pm0.072$ \\
\midrule
\multirow{2}{*}{Our Method} & CTC-TTS-F & 158.47M & 8.02 & 4.20 & \textbf{0.880} & \underline{3.903} & \underline{$4.16\pm0.064$} & $3.85\pm0.071$ \\
& CTC-TTS-L & 159.58M & \textbf{6.33} & \textbf{3.21} & \underline{0.878} & \textbf{3.971} & $\mathbf{4.23\pm0.060}$ & $\mathbf{3.98\pm0.073}$ \\
\midrule
\multirow{3}{*}{Ablation} & CTC+ELLA-V & 159.58M & 20.86 & 11.73 & 0.869 & 3.848 & $3.88\pm0.073$ & \underline{$3.94\pm0.073$} \\
& MFA+ELLA-V & 159.58M & 34.89 & 19.58 & 0.872 & 3.873 & $3.75\pm0.071$ & $3.88\pm0.074$ \\
& MFA+bi-word & 159.58M & \underline{7.53} & \underline{3.99} & 0.874 & 3.840 & $4.14\pm0.068$ & $3.83\pm0.076$ \\
\bottomrule
\end{tabular}
\end{table*}

\subsection{Training configuration}
We use 12 Transformer decoder layers with 16 attention heads, 1024 embedding dimension, 4096 feed-forward dimension, and 0.3 dropout. For CTC-TTS-F, text and speech embedding dimensions are 256 and 768. For single-speaker experiments, the parameters are adjusted to 4 decoder layers, 12 attention heads, 768 embedding dimension, 3072 feed-forward dimension, 0 dropout, 256 text embedding dimension, and 512 speech embedding dimension. All models are trained on 4 RTX 3090 GPUs with a total batch size of 32. We use the AdamW \cite{adamw} optimizer ($\beta_{1}=0.9$, $\beta_{2}=0.95$, $\epsilon=10^{-6}$) and a cosine learning rate scheduler with warm-up to $3\times10^{-4}$ in 25k steps. Training steps are 320k for multi-speaker and 1M for single-speaker. The weight decay is set to 0.1. We use flash-attention \cite{flashattention} for fast training and KV-Cache for inference.

\subsection{Evaluation metrics}
\textbf{Objective metrics}. We use Word Error Rate (WER) and Character Error Rate (CER) to evaluate intelligibility. \texttt{whisper-large-v3} \cite{whisper} and Conformer-Transducer\footnote{\url{https://huggingface.co/nvidia/stt_en_conformer_transducer_xlarge}} are used as ASR models for single- and multi-speaker experiments, respectively. Speech naturalness is measured by UTMOS \cite{utmos}. First-packet latency (FPL-A, assuming the full text is available for TTS) is reported for single-speaker experiments. Speaker similarity (SPK) is computed using \texttt{WavLM-Base-Plus-SV} \cite{wavlm} embeddings. 

\noindent\textbf{Subjective metrics}. We conduct Mean Opinion Score (MOS) for naturalness and Similarity MOS (SMOS) for speaker similarity. 30 samples are randomly selected per system and evaluated by 20 listeners on a 1--5 scale, with results reported as mean scores with 95\% confidence intervals.

\section{Experimental results}
\subsection{Single-speaker streaming experiments}
Table~\ref{tab:ctc_llmvox_compare} shows the results of LLMVox and the two CTC-TTS variants using greedy search. Compared to LLMVox, CTC-TTS-F achieves lower WER and CER while maintaining a shorter FPL-A (ms). The key difference lies in its adoption of CTC alignment and bi-word based sequences, demonstrating the advantages of these two designs. Compared to CTC-TTS-F, CTC-TTS-L delivers even lower WER and CER with a slightly higher FPL-A. The two variants can be selected based on latency and quality requirements. All three methods have identical UTMOS scores (reflecting comparable naturalness), as UTMOS prioritizes speech naturalness over content.

\subsection{Multi-speaker zero-shot experiments}
For the multi-speaker experiments, we reproduce the ELLA-V setup on the zero-shot task using MFA alignment and ELLA-V's training sequence, which we refer to as MFA+ELLA-V. We further conduct ablation studies with different alignment and training sequences: CTC alignment with ELLA-V's sequence (CTC+ELLA-V), and MFA alignment with CTC-TTS-L's sequence (MFA+bi-word). Following ELLA-V, we use nucleus sampling in multi-speaker generation for more stable results.

\textbf{Continuation}. Given a text segment and its corresponding 3-second prefixed speech, the task aims to synthesize speech for the remaining text. Results are shown in Table~\ref{tab:continuation}. For methods using bi-word sequences, CTC-TTS-F achieves competitive performance compared with MFA+bi-word, while CTC-TTS-L outperforms MFA+bi-word in all metrics except SPK---demonstrating the superiority of CTC alignment. For methods using CTC alignment, our methods outperform CTC+ELLA-V across the board, which may stem from the limited context provided by ELLA-V's local advance mechanism and further validates the advantage of the bi-word interleaving scheme. Notably, CTC-TTS-L delivers near-optimal performance on the entire continuation task, highlighting the merits of combining CTC alignment, bi-word training sequences, and length-wise token concatenation.

\textbf{Cross-speaker.} Given about 3 seconds of speech and its transcribed text as a prompt, this task synthesizes speech for another utterance. Results are shown in Table~\ref{tab:cross}. Due to the out-of-domain testing, all results are slightly worse than in the continuation task, but the overall trends are similar. As in the continuation task, CTC-TTS-L achieves near-optimal performance on the cross-speaker task. CTC-TTS-F is slightly worse than MFA+bi-word; however, CTC-TTS-F uses feature-level stacking, whereas MFA+bi-word uses length-wise concatenation, so they are not directly comparable. 
Overall, combining CTC alignment with bi-word sequences achieves strong performance in this out-of-domain setting.
When using ELLA-V's training sequence, both CTC+ELLA-V and MFA+ELLA-V yield poor performance, highlighting the limitations of ELLA-V's sequence organization. Notably, CTC+ELLA-V outperforms MFA+ELLA-V in WER and CER, whereas MFA+ELLA-V is better on the in-domain continuation task. 
This indicates that MFA alignment performs well for in-domain continuation, while CTC alignment generalizes better to out-of-domain cross-speaker scenarios.

\section{Conclusion}
This paper presents CTC-TTS, an LLM-based dual-streaming TTS method that derives phoneme–speech correspondence via CTC alignment and trains on bi-word interleaved sequences. Two variants, CTC-TTS-L and CTC-TTS-F, implement bi-word interleaving along the length and feature dimensions respectively, offering a practical quality–latency trade-off. Experiments show that CTC-TTS outperforms fixed-ratio interleaving and MFA-based baselines on streaming and zero-shot tasks. Future work includes adopting neural G2P models such as the CTC-based one in JSA-SPG \cite{jsaspg}, and leveraging neural forced alignment to obtain more precise boundaries.

\section{Generative AI Use Disclosure}
Generative AI tools are used in this work only for language editing, polishing, and formatting of the manuscript. They are not used to generate any core content, research ideas, experimental designs, results, or major textual parts of the paper. All scientific contributions, including model design, experiments, analysis, and conclusions, are completed by the authors.

\section{Acknowledgements}
This work was supported by the National Natural Science Foundation of China (Grant No. 62466055)

\bibliographystyle{IEEEtran}
\bibliography{mybib}

\end{document}